\documentclass[aps,prl,twocolumn,amsmath,longbibliography]{revtex4-2}
\usepackage{graphicx}
\usepackage{subfigure}
\usepackage{url}
\usepackage{dcolumn}
\usepackage{bm}
\usepackage[]{natbib}
\usepackage{mathrsfs}
\usepackage{amsmath}
\usepackage{amsfonts}
\usepackage[colorlinks=true, 
citecolor=blue, 
anchorcolor=blue, 
linkcolor=blue, 
urlcolor=black]{hyperref}
\usepackage{graphicx,epstopdf}
\usepackage{subfigure}
\usepackage{epsfig}
\usepackage{dcolumn}
\usepackage{bm}
\usepackage{color}
\usepackage{natbib}
\usepackage{amssymb}
\usepackage{xcolor}
\usepackage{braket}
\begin{document}

\title{Higher-order Angular Momentum Transport of Alteraxial Phonons in 2D Magnets}

\author{Linxin Liu$^{1}$, Fuyi Wang$^{2}$, and Huaiqiang Wang$^{1,3,\ast}$}

\affiliation{
$^1$ Center for Quantum Transport and Thermal Energy Science, Institute of Physics Frontiers and Interdisciplinary Sciences, School of Physics and Technology, Nanjing Normal University, Nanjing 210023, China\\
$^2$ National Laboratory of Solid State Microstructures and School of Physics, Nanjing University, Nanjing 210093, China\\
$^3$ Jiangsu Physical Science Research Center, Nanjing 210093, China}

\email{hqwang@njnu.edu.cn}

\begin{abstract}
The coupling between lattice vibrations and magnetic order in two-dimensional (2D) magnets gives rise to a variety of physical phenomena, among which magnetic effects of axial phonons carrying phonon angular momentum (PAM) have attracted considerable interest. However, a systematic classification of axial phonons in 2D magnets that covers their multipolar patterns and the resulting transport manifestations remains largely unexplored. Here, based on symmetry analysis, we classify all 2D magnetic point groups by their long-wavelength PAM wave patterns.  Beyond the $s$-wave axial phonon, we identify alteraxial phonons with \(p\)-, \(d\)-, \(f\)-, \(g\)-, and \(i\)-wave multipolar textures, characterized by 1, 2, 3, 4, and 6 zero-PAM nodal lines, respectively. We then develop a theory of nonlinear PAM transport driven by a temperature gradient and derive two selection rules: a parity selection rule that determines whether the response occurs at even or odd orders, and an angular-harmonic selection rule that fixes the lowest allowed order for each wave pattern. The multipolar PAM patterns and selection rules are exemplified by two effective lattice models realizing \(f\)-wave and \(d\)-wave PAM patterns, respectivley. The angular dependence of the nonlinear PAM current reflects the underlying multipolar PAM texture, offering a signature for experiments. Our results establish a symmetry-guided framework for detecting and utilizing higher-order PAM textures in 2D magnets.

\end{abstract}

\maketitle

\noindent\emph{Introduction.}  The interplay between lattice vibrations and magnetism, as manifested in spin-phonon and magnon-phonon interactions, has long been an important topic in condensed matter physics. Beyond energy and momentum, phonons can also carry angular momentum arising from rotational lattice vibrations \cite{zhang_angular_2014}. As an additional degree of freedom, phonon angular momentum (PAM) underlies the concept of phonon chirality, which has attracted considerable interest recently~\cite{zhang_chiral_2015, Sun2026, zhu_observation_2018, ueda_chiral_2023, basini_terahertz_2024, chen_entanglement_2019, grissonnanche_chiral_2020, ishito_truly_2023,  kahana_light_2024, klebl_ultrafast_2025, chen_chiral_2018,  paiva_dynamically_2025,  hernandez_chiral_2023, kim_chiral-phonon-activated_2023, luo_transverse_2023, hernandez_observation_2023, romao_phonon-induced_2024, Coh2023Classification, chen_emergence_2025, zhang_advances_2025, suzuki_disorder_2025, chen2022chiral, nabei_orbital_2026, ren_phonon_2021, zhang_gate-tunable_2023, luo2023nonlinear, gao2023chiral, mustafa_origin_2025, xue_extrinsic_2025, Hamada2018, park_phonon_2020, saparov_lattice_2022, zhang2025measurement, merlin_magnetophononics_2024, liu_collective_2026, pols2026chiral, wang2026inelastic, wang2026catalog, chen2026nonadiabatic}. Phonons with nonzero PAM, termed axial phonons~\cite{juraschek_chiral_2025}, can couple to external magnetic fields or internal magnetic moments, giving rise to a variety of phono-magnetic phenomena~\cite{davies_phononic_2024, yin_chiral_2021, liu_model_2017, juraschek_orbital_2019, flebus_phonon_2023, bonini_frequency_2023,  tauchert_polarized_2022, lujan_spinorbit_2024, shabala_phonon_2024, ren_adiabatic_2024, bowen_chiral_2024, okamoto_altermagnetic_2025, shabala_axial_2025, che_magnetic_2025, yang_inherent_2025, zhang_general_2026, zhang_comprehensive_2026, Yarmohammadi_2026, Bao2026, thingstad_chiral_2019, nomura_phonon_2019, wu_fluctuation-enhanced_2023,  wu_magnetic_2025, chatterjee2026phonon, wang_magnetic_2024, weissenhofer_chiral_2025} such as the phonon Hall effect~\cite{zhang2011phonon, qin_berry_2012}, the phonon Zeeman effect~\cite{juraschek_giant_2022, wang_ab_2025}, and the ultrafast Barnett effect~\cite{davies_phononic_2024}.

Analogous to altermagnets, which exhibit momentum-dependent spin splitting and multipolar spin textures~\cite{smejkal_emerging_2022, smejkal_beyond_2022, song_altermagnets_2025}, recent studies have shown that PAM can display not only conventional \(s\)-wave patterns but also higher-order wave patterns in momentum space~\cite{2026Bendin, chatzichrysafis2026angular, leon2026chiral, rieger2026symmetry, chatzichrysafis2026angular, you2026structural, yang2026symmetry, wang2025alteraxial}. For instance, a \(d\)-wave PAM texture and the associated splitter effect have been proposed in altermagnets via magnon--phonon hybridization~\cite{2026Bendin, chatzichrysafis2026angular}. Moreover, Yang \textit{et al.} classified phonon chirality using crystallographic space groups and identified phonons with higher-order helicities \cite{yang2026symmetry}, while Wang \textit{et al.} categorized PAM patterns in collinear magnets via magnetic point groups (MPGs) and introduced the concept of alteraxial phonons~\cite{wang2025alteraxial}. Nevertheless, a systematic symmetry-based investigation of axial phonons in two-dimensional (2D) magnets remains lacking. Given the rapid growth of 2D magnets, a symmetry-guided framework for axial phonons that covers their multipolar patterns and the resulting transport manifestations is highly desirable.

In this work, we provide a complete symmetry classification of long-wavelength phonons in all 2D MPGs, identifying a series of alteraxial phonons with \(p\)-, \(d\)-, \(f\)-, \(g\)-, and \(i\)-wave PAM textures beyond the conventional \(s\)-wave axial phonon. We then formulate a nonlinear theory of PAM transport under a temperature gradient and derive two selection rules, namely, a parity selection rule and an angular-harmonic selection rule, which govern the order and angular dependence of the PAM response. To illustrate these principles, we construct two effective lattice models realizing the \(3m.1'\) and \(4'\) MPGs with \(f\)-wave and \(d\)-wave PAM patterns, respectively, and show that the nonlinear PAM current exhibits characteristic angular oscillations that directly reflect the underlying multipolar texture. Our results establish a symmetry-guided framework for exploring higher-order PAM transport in 2D magnets.\\

\begin{figure*}[t]
    \centering
   \includegraphics[width=0.9\linewidth]{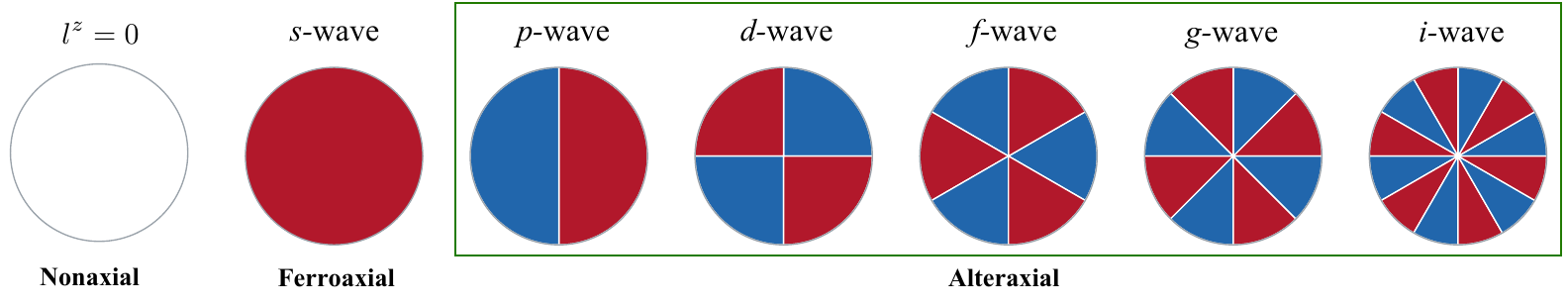}
    \caption{Schematic illustration of phonon angular momentum (PAM) patterns in 2D magnets. Red and blue colors denote opposite signs of the out-of-plane (\(z\)) component of the PAM. In contrast to nonaxial phonons carrying no PAM, The axial phonons exhibit \(s\)-, \(p\)-, \(d\)-, \(f\)-, \(g\)-, and \(i\)-wave PAM patterns, with 0, 1, 2, 3, 4, and 6 symmetry-guaranteed zero-PAM nodal lines, respectively. Phonons with higher-order PAM patterns are termed alteraxial phonons.}
    \label{Fig.1}
\end{figure*}

\noindent\emph{PAM classification for 2D magnets.} For two-dimensional (2D) magnets, the out-of-plane \(z\) direction is naturally chosen as the quantization axis of PAM. For a phonon eigenmode \(\epsilon_{\mathbf{q}}\) at wavevector \(\mathbf{q}=(q_x,q_y)\), the \(z\) component of the PAM is obtained from the phonon circular polarization as \cite{zhang_chiral_2015}
\begin{equation}
	\label{jz}
	l^{z}_\mathbf{q}=\langle\epsilon_{\mathbf{q}}|\left(|R\rangle\langle R|-|L\rangle\langle L|\right)|\epsilon_{\mathbf{q}}\rangle\hbar.
\end{equation}
Here, \(\hbar\) is the reduced Planck constant, and \(|R\rangle\) (\(|L\rangle\)) denotes the right-handed (left-handed) circularly polarized eigenstate with PAM \(+\hbar\) (\(-\hbar\)) along the \(z\) direction. Since phonons arise from real-space lattice vibrations, their coupling to magnetic order necessitates the use of magnetic groups with coupled real and spin spaces. In this work, we focus on long-wavelength \(\Gamma\) phonons at the Brillouin-zone center of 2D magnets, whose symmetry is governed by 2D MPGs, denoted by \(G\). For a symmetry operation \(\mathcal{G}\in G\), the PAM \(l^z_\mathbf{q}\) must satisfy \cite{wang2025alteraxial}
\begin{equation}
	\label{sym}
	l^z_{\mathcal{G}\mathbf{q}}=\eta(\mathcal{T})\chi(\mathcal{R})l^z_{\mathbf{q}},
\end{equation}
where \(\mathcal{T}\) and \(\mathcal{R}\) denote the time-reversal and point-group parts of \(\mathcal{G}\), respectively. Since \(\mathcal{T}\) reverses the sign of the PAM, the corresponding coefficient is \(\eta(\mathcal{T})=-1\) if \(\mathcal{T}\) is contained in \(\mathcal{G}\), and \(\eta(\mathcal{T})=+1\) otherwise. Similarly, if \(\mathcal{R}\) reverses (preserves) \(l^z\), then \(\chi(\mathcal{R})=-1\) (\(+1\)). Consequently, when the product \(\eta(\mathcal{T})\chi(\mathcal{R})\) equals \(-1\), \(l^z\) changes sign between the phonon modes at \(\mathbf{q}\) and \(\mathcal{G}\mathbf{q}\).

In the presence of inversion symmetry \(\mathcal{P}\) (note that in 2D MPGs, inversion is equivalent to the twofold rotation \(C_{2z}\)), we have \(l^{z}_{-\mathbf{q}}=l^z_{\mathbf{q}}\), indicating that \(l^{z}\) is even under \(\mathcal{P}\). In contrast, \(l^{z}\) is odd under time reversal \(\mathcal{T}\), with  \(l^{z}_{-\mathbf{q}}=-l^z_{\mathbf{q}}\). As a result, the PAM \(l^{z}\) is forced to vanish when a combined \(\mathcal{PT}\) symmetry exists in a 2D MPG. This corresponds to nonaxial phonons in 10 MPGs for 2D magnets (see Table I). Without the combined \(\mathcal{PT}\) symmetry, a nonzero \(l^z_{\mathbf{q}}\) is generically allowed, leading to the emergence of axial phonons. Specifically, if \(G\) contains at least one operation satisfying \(\eta(\mathcal{T})\chi(\mathcal{R})=-1\), \(l^z\) is forced to vanish at \(\Gamma\) and alternates in sign in the surrounding Brillouin zone, thereby yielding alteraxial phonons. This condition can be satisfied by two types of operations: one with \(\chi(\mathcal{R})=-1\) and \(\eta(\mathcal{T})=+1\), and the other with \(\chi(\mathcal{R})=+1\) and \(\eta(\mathcal{T})=-1\). In 2D MPGs, only vertical mirror operations \(\sigma_v\) satisfy the first condition. Since \(\sigma_v \mathbf{q}=\mathbf{q}\) for wavevectors lying on the mirror-invariant line, \(l^z\) must vanish along this line, giving rise to a symmetry-enforced nodal line crossing $\Gamma$ with vanishing PAM. As for the second condition, it is satisfied by the time-reversal operation or its combination with $C_{3z}$ rotation (other $C_{nz}$ rotations would enforce the combined $\mathcal{PT}$ symmetry in 2D). This gives rise to symmetry-protected zero-PAM nodal lines enclosing \(\Gamma\), and unlike the preceding mirror-invariant line, they are not pinned to any high-symmetry lines. Moreover, if \(G\) contains no symmetry operation yielding \(\eta(\mathcal{T})\chi(\mathcal{R})=-1\), a nonzero \(l^z\) is allowed at \(\Gamma\), and no symmetry-driven zero-PAM nodal lines cross \(\Gamma\). 

The total number of zero-PAM nodal lines passing through $\Gamma$ is determined from the symmetry constraint in Eq. (\ref{sym}) and is  termed the MPG integer $n_{\mbox{\tiny {MPG}}}^{\mbox{\tiny{2D}}}$~\cite{wang2025alteraxial}. A comprhensive analysis of all 31 2D MPGs (10 gray, 10 white, and 11 black-and-white) reveals that $n_{\mbox{\tiny {MPG}}}^{\mbox{\tiny{2D}}}$ takes the values \(0, 1, 2, 3, 4\), and \(6\) (see Table I for the complete classification). These correspond to \(s\)-, \(p\)-, \(d\)-, \(f\)-, \(g\)-, and \(i\)-wave PAM patterns in the 2D Brillouin zone, respectively, as illustrated in Fig.~\ref{Fig.1}, covering both odd and even waves. The phonons exhibiting higher-order-wave PAM patterns are referred to as alteraxial phonons~\cite{wang2025alteraxial}.\\

\begin{table*}[t]
	\centering
	\caption{Complete classification of the 31 (10 gray, 10 white, and 11 black-and-white) two-dimensional magnetic point groups (MPGs, given in UNI notation~\cite{campbell_introducing_2022}) based on distinct PAM wave patterns according to the MPG integer \(n_{\mathrm{MPG}}^{2\mathrm{D}}\) defined as the symmetry-guranteed zero-PAM nodal lines.For each PAM pattern, the representative polynomial(where \(x\) and \(y\) are interchangeable) and the order of the PAM current response \(J^{l_z(N)}\) under an in-plane temperature gradient are also provided. }
	\label{tab:mpg_classification}
	\small
	\setlength{\tabcolsep}{4pt}
	\renewcommand{\arraystretch}{1.3}
	\begin{tabular}{c c c c c}
		\hline\hline
		PAM-wave & 2D MPG & Representative Polynomial & \(n_{\mathrm{MPG}}^{2\mathrm{D}}\) & \(J^{l_z(N)}\) \\
		\hline
		Nonaxial & \(2'.1'\), \(2'\), \(4'.1'\), \(6'.1'\), \(6'\), \(mm2\,.1'\), \(m'm2'\), \(4mm\,.1'\), \(6mm\,.1'\), \(6'm'm\) & -- & -- & -- \\
		\(s\) & \(1\), \(2\), \(3\), \(4\), \(6\), \(m'\), \(m'm'2'\), \(3m'\), \(4m'm'\), \(6m'm'\) & \(1\) & \(0\) & 1,3,5,... \\
		\(p\) & \(1\,.1'\), \(m\), \(m\,.1'\) & \(q_y\) & \(1\) & 2,4,6,... \\
		\(d\) & \(4'\), \(mm2\), \(4'm'm\) & \(q_x q_y\) & \(2\) & 1,3,5,... \\
		\(f\) & \(3\,.1'\), \(3m\), \(3m\,.1'\) & \(q_y(3q_x^2-q_y^2)\) & \(3\) & 2,4,6,... \\
		\(g\) & \(4mm\) & \(q_x q_y(q_x^2-q_y^2)\) & \(4\) & 3,5,7,... \\
		\(i\) & \(6mm\) & \(q_x q_y(3q_x^4-10q_x^2 q_y^2+3q_y^4)\) & \(6\) & 5,7,9,... \\
		\hline\hline
	\end{tabular}
\end{table*}

\noindent\emph{Higher-order PAM transport.} The existence of higher-order PAM patterns in alteraxial phonons gives rise to a unique multipolar higher-order PAM current response under an external temperature gradient, as we demonstrate below. We consider a 2D phonon system, in which the phonon frequency and the \(z\) component of the PAM at wavevector \(\mathbf{q}\) for the \(n\)-th phonon band are denoted by \(\omega_{n\mathbf{q}}\) and \(l^z_{n\mathbf{q}}\), respectively. The corresponding phonon velocity is given by \(\mathbf{v}_{n\mathbf{q}}=\nabla_\mathbf{q}\omega_{n\mathbf{q}}\). At equilibrium, the phonon occupation is given by the Bose-Einstein distribution \(f_{n\mathbf{q}}^0 = 1/[e^{\hbar\omega_{n\mathbf{q}}/(k_B T)} - 1]\) with $T$ denoting the temperature.  In the presence of an external temperature gradient, the nonequilibrium PAM response current can be written as~\cite{2026Bendin}
\begin{equation}
	\label{PAM current}
  \mathbf{J}^{l^z}=\sum_n\mathbf{\int_q}l^{z}_{n\mathbf{q}}\mathbf{v}_{n\mathbf{q}}\delta f_{n\mathbf{q}},
\end{equation}
where $\delta f_{n\mathbf{q}}= f_{n\mathbf{q}}-f_{n\mathbf{q}}^0$. Here, assuming no other external perturbations and a slowly varying temperature gradient, the steady-state phonon Boltzmann transport equation in the relaxation-time approximation reads ~\cite{battiato_boltzmann_2017, nakai_nonreciprocal_2019}
\begin{equation}
	\mathbf{v}_{n\mathbf{q}}\cdot\nabla_{\mathbf{r}}f_{n\mathbf{q}}
	= -\frac{f_{n\mathbf{q}}-f_{n\mathbf{q}}^0}{\tau_{n\mathbf{q}}},
\end{equation}
where \(\tau_{n\mathbf{q}}\) is the relaxation time. For notational brevity, we suppress the subscript \(n\mathbf{q}\) unless otherwise specified. Introducing the temperature gradient as the driving force, \(\mathbf{F}=-\nabla_{\mathbf{r}}T\), the distribution function is expanded as
\(f=f^0+\sum_{N=1}^{\infty}f^{(N)}\),
with \(f^{(N)}=O(F^N)\) satisfying the recurrence relation
\(f^{(N)}=-\tau\,\mathbf{v}\cdot\nabla_{\mathbf{r}}f^{(N-1)}\).
Using \(\mathbf{v}\cdot\nabla_{\mathbf{r}}=-(\mathbf{v}\cdot\mathbf{F})\partial_T\), we obtain
\(f^{(N)}=\tau^N(\mathbf{v}\cdot\mathbf{F})^N\partial_T^N f^0\).
Accordingly, the \(N\)-th order PAM current response is given by
\begin{equation}
	\label{N-th PAM current}
	J_\alpha^{l^z (N)}
	= \chi_{\alpha \beta_1\cdots \beta_N}^{(N)}
	F_{\beta_1}\cdots F_{\beta_N},
\end{equation}
where \(\alpha\) and \(\beta_i\) label Cartesian components, and the response coefficient is
\begin{equation}
	\chi_{\alpha \beta_1\cdots \beta_N}^{(N)}
	= \sum_n \int_{\mathbf{q}}
	\tau^N l^z v^\alpha v^{\beta_1}\cdots v^{\beta_N}
	\partial_T^N f^0.
\end{equation}

We show that the PAM wave patterns and parities of 2D alteraxial phonons impose distinct selection rules on the higher-order PAM current: the parity selection rule, which determines whether the response is of even or odd order, and the angular-harmonic selection rule, which determines the lowest-order response. For the first rule, when either inversion (\(C_{2z}\)) or time-reversal symmetry is present, as is the case for most alteraxial phonons in 2D magnets (only the MPGs $m$ and $3m$ are exceptions, see Table I), we have \(\omega_{n\mathbf{q}}=\omega_{n-\mathbf{q}}\) and \(l^z_{n\mathbf{q}}=\xi\,l^z_{n-\mathbf{q}}\), where \(\xi=+1\) for even-wave patterns and \(\xi=-1\) for odd-wave patterns. Consequently, the velocity is odd under \(\mathbf{q}\to-\mathbf{q}\), i.e., \(v_{n\mathbf{q}}=-v_{n-\mathbf{q}}\). Assuming also \(\tau_{n\mathbf{q}}=\tau_{n-\mathbf{q}}\), the response coefficient \(\chi^{(N)}\), which contains one PAM factor and \(N+1\) velocity factors, vanishes unless \(\xi(-1)^{N+1}=1\). Therefore, alteraxial phonons with odd-parity (even-parity) PAM patterns exhibit only even-order (odd-order) PAM current responses. 

For the second rule, we simplify the analysis by considering an isotropic in-plane phonon dispersion, \(\omega_{n\mathbf{q}}=\omega_{n}(q)\), and an isotropic relaxation time, \(\tau_{n\mathbf{q}}=\tau_{n}(q)\), where \(q=|\mathbf{q}|\). In this case, the velocity \(\mathbf{v}_{n\mathbf{q}}\) is parallel to \(\mathbf{q}\), so that \(\mathbf{v}_{n\mathbf{q}}=v_{n}(q)\hat{e}_{\mathbf{q}}=v_{n}(q) (\cos\theta, \sin\theta)\). We further assume that the 2D PAM pattern consists of a single angular harmonic, \(l_{n}^{z}=l_{nl}(q)\cos(l\theta)\), where \(l=1,2,3,4,6\) labels the \(p\), \(d\), \(f\), \(g\), and \(i\) wave orders, respectively. In addition, we apply a temperature gradient along the unit vector \(\hat{e}_{\parallel}=(\cos\phi,\sin\phi)\), with the driving force given by \(\mathbf{F}=F\hat{e}_{\parallel}\).  The \(N\)-th order longitudinal PAM current response, which is obtained from Eq.~(\ref{N-th PAM current}) as (see the Supplemental Material~\cite{supp} for details.)
\begin{equation}
	\label{J PAM}
	J_\parallel^{l^z (N)}=\mathcal{R}_{Nl}B_{Nl}F^N\cos(l\phi).
\end{equation}
Here, $\mathcal{R}_{Nl}$ and $B_{Nl}$ arise from radial and angular integrations, respectively, and are given by
\begin{align}
	\mathcal{R}_{Nl} &= \sum_n\int_0^{q_c}\frac{q\,dq}{(2\pi)^2}\,
	\tau_n^N(q)\,l^z_{nl}(q)\,v_n^{N+1}(q)\,\partial_T^N f^0,
	\label{RNl}\\
	B_{Nl} &=
	\begin{cases}
		\displaystyle \frac{\pi}{2^N}\binom{N+1}{\frac{N+1-l}{2}}, & N+1\ge l,\; \frac{N+1-l}{2}\in \mathbb{Z},\\[6pt]
		0, & \text{otherwise}.
	\end{cases}
	\label{BNl}
\end{align} 
Notably, \(B_{Nl}\) is nonzero only when \(N+1\ge l\) and \(N+1-l\) is even, thereby imposing the angular harmonic selection rule. Consequently, for an \(l\)-wave PAM pattern, the lowest allowed order of the longitudinal PAM current response is \(N_{\mbox{\tiny{min}}}=l-1\) for \(l\ge2\), while for \(l=1\) ($p$-wave) it is \(N_{\mbox{\tiny{min}}}=2\) (see the SM~\cite{supp} for detailed expressions for the lowest-order response of each wave).

We now remark on the validity of the selection rules. High-symmetry 2D magnets exhibit nearly isotropic in-plane phonon branches near \(\Gamma\), whereas low-symmetry MPGs can be anisotropic. Thus, our isotropic-dispersion assumption serves as a simplifying limit that captures the essential angular-harmonic selection rules, whereas a quantitative treatment of low-symmetry cases would require the full anisotropic sound-velocity tensor.\\

\begin{figure}[!t]
	\centering
	\includegraphics[width=3.4in]{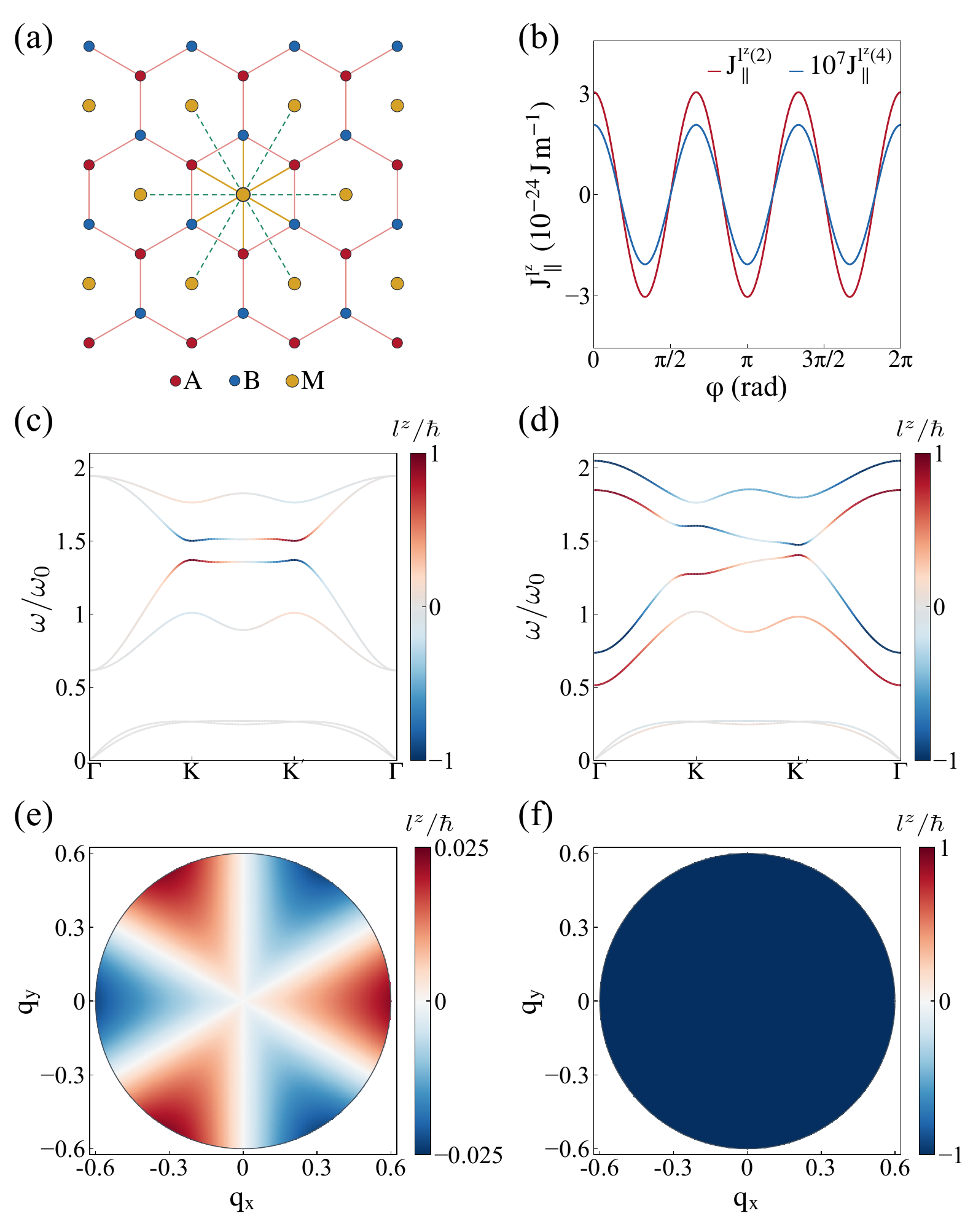}
	\caption{(a) Schematic of the 2D lattice model hosting the \(f\)-wave PAM. The model consists of a honeycomb lattice with two inequivalent nonmagnetic atoms (denoted by A and B) and a triangular lattice of magnetic atoms (denoted by M). Solid lines denote nearest-neighbor bonds between  A, B, and M atoms, and green dashed lines represent bonds between nearest-neighbor M atoms. For the paramagnetic case (MPG \(3m.1'\)): (b) the second- and fourth-order longitudinal PAM response currents versus the in-plane temperature-gradient direction, (c) representative phonon bands, and (e) PAM distributions around \(\Gamma\). For the ferromagnetic case with out-of-plane moments (MPG \(3m'\)): (d) representative phonon bands and (f) PAM distributions around \(\Gamma\). Phonon frequencies are given in units of \(\omega_0\) and are therefore dimensionless. Details of the numerical parameters are provided in the Supplemental Material~\cite{supp} .}
	\label{Fig.2}
\end{figure}

\noindent\emph{Effective lattice-model realizations.} To illustrate alteraxial phonons and PAM transport in 2D magnets, we construct two effective models realizing the \(3m.1'\) and \(4'\) MPGs, with long-wavelength phonons exhibiting odd-parity \(f\)-wave and even-parity \(d\)-wave PAM patterns, respectively. The 2D MPG \(3m.1'\) can be realized in the paramagnetic phase of a honeycomb lattice with two inequivalent atoms of different masses and a triangular lattice with a magnetic atom located at the center of each honeycomb, as illustrated in Fig.~\ref{Fig.2}(a). For simplicity, we assume a heavy magnetic atom so that its phonon bands are separated from the others. Figure~\ref{Fig.2}(c) shows typical phonon bands in this phase, with the corresponding PAM explicitly labeled (see the SM for detailed calculations). Here, PAM arises from the broken inversion symmetry due to the two inequivalent nonmagnetic atoms, while time-reversal symmetry forces it to vanish at \(\Gamma\). As shown in Fig.~\ref{Fig.2}(e), the PAM around \(\Gamma\) exhibits an \(f\)-wave-like distribution that vanishes on the three \(C_{3z}\)-connected mirror lines, consistent with our classification.  Under an external temperature gradient, such an \(f\)-wave PAM yields even-order longitudinal PAM current responses (starting from second order), which exhibit three periodic sinusoidal oscillations as the in-plane temperature-gradient direction is rotated, as illustrated by the second- and fourth-order responses in Fig.~\ref{Fig.2}(b). Interestingly, if the system enters a ferromagnetic phase with out-of-plane magnetic moments, the 2D MPG changes from \(3m.1'\) to \(3m'\). Consequently, the PAM pattern changes from \(f\)-wave to \(s\)-wave: the phonon bands at \(\Gamma\) split into states with opposite PAM [see Fig.~\ref{Fig.2}(d)], and the PAM no longer changes sign around \(\Gamma\), as shown by its distribution in Fig.~\ref{Fig.2}(f).

\begin{figure}[!t]
	\centering
	\includegraphics[width=3.2in]{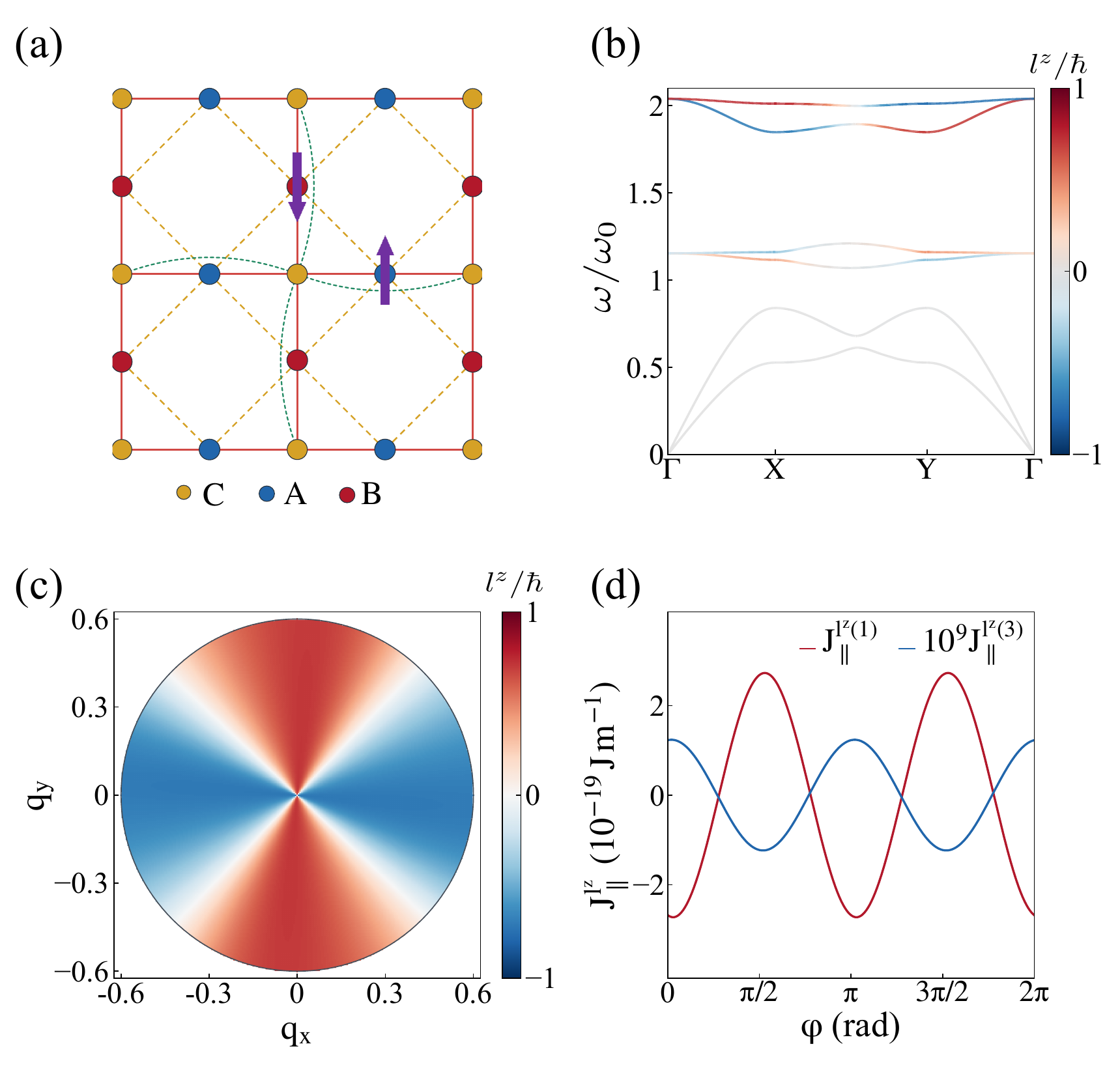}
	\caption{(a) Schematic of the 2D Lieb lattice model with MPG \(4'\), hosting the \(d\)-wave PAM. The lattice consists of one nonmagnetic atom (C) and two \(C_{4z}\mathcal{T}\)-connected magnetic atoms (A and B) with opposite out-of-plane magnetic moments. Red solid lines denote nearest-neighbor bonds, yellow dashed lines denote next-nearest-neighbor bonds between A and B, and green dashed lines denote bonds between adjacent C atoms. (b) Representative phonon bands, (c) PAM distributions around \(\Gamma\), and (d) the first- and third-order longitudinal PAM response currents versus the in-plane temperature-gradient direction. Phonon frequencies are given in units of \(\omega_0\) and are therefore dimensionless. Details of the numerical parameters are provided in the Supplemental Material~\cite{supp} .}
	\label{Fig.3}
\end{figure}

To demonstrate even-parity \(d\)-wave alteraxial phonons in 2D magnets, we propose a Lieb lattice governed by the MPG \(4'\) [Fig.~\ref{Fig.3}(a)], comprising one nonmagnetic atom and two identical magnetic atoms with opposite out-of-plane moments related by \(C_{4z}\mathcal{T}\). Figures~\ref{Fig.3}(b) and~\ref{Fig.3}(c) show the phonon bands and PAM distribution around \(\Gamma\), where the PAM vanishes at \(\Gamma\) and exhibits a \(d\)-wave-like pattern with two zero-PAM nodal lines. This even-parity \(d\)-wave (\(l=2\)) state, with preserved \(\mathcal{P}\) (\(C_{2z}\)) and broken \(\mathcal{T}\), produces odd-order (\(1,3,5,\dots\)) PAM current responses under a temperature gradient, as shown in Fig.~\ref{Fig.3}(d). The response magnitude oscillates twice as the in-plane temperature-gradient direction is rotated, which is consitent with our theory and provides a method to detect the $d$-wave alteraxial phonons.\\

\noindent\emph{Summary.} In summary, we have presented a comprehensive symmetry-based study of PAM patterns in 2D magnets. A complete classification of PAM wave patterns for all 31 MPGs is established, uncovering a hierarchy of alteraxial phonons with \(p\)-, \(d\)-, \(f\)-, \(g\)-, and \(i\)-wave textures. We further developed a nonlinear transport theory for PAM under a temperature gradient, from which two selection rules emerge: the parity selection rule and the angular-harmonic selection rule. These rules dictate the allowed orders and angular symmetries of the nonlinear PAM current. Two minimal lattice models, a honeycomb lattice for the \(3m.1'\) group with \(f\)-wave PAM and a Lieb lattice for the \(4'\) group with \(d\)-wave PAM, serve as concrete realizations. The computed angular dependence of the nonlinear current matches the symmetry analysis, offering a practical way to identify multipolar PAM textures in experiments. Our findings open an avenue for exploiting nonlinear phonon angular momentum transport in 2D magnets.\\

\begin{acknowledgments}
This work is supported by the National Natural Science Foundation of China (Grant No. 12534007) and the Natural Science Foundation of Jiangsu Province (Grant Nos. BK20252117 and BK20233001).

L. L. and F. W. contributed equally to this work.
\end{acknowledgments}

\bibliography{references}
\appendix

\end{document}